\documentclass[twocolumn]{svjour3}

\smartqed

\usepackage{cite}
\usepackage{color}
\usepackage{graphicx}
\usepackage{epstopdf}
\usepackage{amsmath,amssymb,amsfonts}
\usepackage{textcomp}
\usepackage{dblfloatfix}
\usepackage[export]{adjustbox}
\usepackage{multirow}
\usepackage{booktabs}
\usepackage{pifont}
\usepackage{epsfig}

\newcommand{\numofinputs}{K}
\newcommand{\numofoutputs}{H}

\newcommand{\startupdelay}{d}
\newcommand{\numofbits}{n}
\newcommand{\clockrate}{c}

\newcommand{\videoin}{V}
\newcommand{\videoout}{U}
\newcommand{\videoindig}{\hat{V}}
\newcommand{\videoindigsync}{\hat{V}^{\text{S}}}

\newcommand{\cmark}{\ding{51}}
\newcommand{\xmark}{\ding{55}}

\newcommand{\Archadj}{Flexible}
\newcommand{\archadj}{flexible}

\begin{document}

\title{{\Archadj} Architecture for Real-time Processing of Multiple Video Signals}

\author{Mohamed Awad         	\and
		Islam T. Abougindia 	\and
		Ahmed Elliethy			\and
		Hussein A. Aly
}

\institute{Mohamed Awad \at
	Military Technical College, Cairo, Egypt \\
	\email{awad.mrms@gmail.com}
	\and
	Islam T. Abougindia \at
	Military Technical College, Cairo, Egypt \\
	\email{iabougindia@ieee.org}
	\and
	Ahmed Elliethy \at
	Military Technical College, Cairo, Egypt \\
	\email{a.s.elliethy@mtc.edu.eg}
	\and
	Hussein A. Aly \at
	Military Technical College, Cairo, Egypt \\
	\email{haly@ieee.org}
}

\date{Received: date / Accepted: date}

\maketitle

\begin{abstract}

Simultaneous processing of multiple video sources requires each pixel in a frame from a video source to be processed synchronously with the pixels at the same spatial positions in corresponding frames from the other video sources. However, simultaneous processing is challenging as corresponding frames from different video signals provided by multiple sources have time-varying delay because of the electrical and mechanical restrictions inside the video sources hardware that cause deviation in the corresponding frame rates. Researchers overcome the aforementioned challenges either by utilizing ready-made video processing systems or designing and implementing a custom system tailored to their specific application. These video processing systems lack flexibility in handling different applications requirements such as the required number of video sources and outputs, video standards, or frame rates of the input/output videos. In this paper, we present a design for a {\archadj} simultaneous video processing architecture that is suitable for various applications. The proposed architecture is upgradeable to deal with multiple video standards, scalable to process/produce a variable number of input/output videos, and compatible with most video processors. Moreover, we present in details the analog/digital mixed-signals and power distribution considerations used in designing the proposed architecture. As a case study application of the proposed {\archadj} architecture, we utilized the architecture for a realization of a simultaneous video processing system that performs video fusion from visible and near-infrared video sources in real time. We make available the source files of the hardware design along with the bill of material (BOM) of the case study to be a reference for researchers who intend to design and implement simultaneous multi-video processing systems.

\keywords{multi-video simultaneous processing \and multi-video synchronization \and real-time video fusion}

\end{abstract}

\section{Introduction}
\label{sec:intro}

Simultaneous processing of multiple video sources has been the core of many state-of-the-art vision applications such as multi-view 3D-Televisions~\cite{fouad:2016:modified,park:2012:realistic,ahmad:2007:multi}, augmented reality~\cite{elliethy:2014:view,van:2010:survey} and fusion-based video enhancement~\cite{ancuti:2012:enhancing,bennett:2007:multispectral}. These applications require simultaneous processing of multiple video sources where each pixel in a frame from a video source is processed synchronously with the pixels at the same spatial positions in the corresponding frames from the other video sources~\cite{bennett:2007:multispectral}.

\begin{table*}[t]
	\centering
	\resizebox{0.97\textwidth}{!}{
		\begin{tabular}{@{}lccccccccc@{}}
			\toprule
			Architecture & scalability & number of & number of & video & different & different & startup & deviation in & independence of \\
			&& input & output & format & spatial & temporal & delay & temporal & output format \\
			&& videos & videos && resolutions & resolutions && resolution & from input \\ \midrule
			Custom~\cite{guo:2017:fpga} & \xmark & 9 & 1 & 4K-UHD & \cmark & \cmark & \xmark & \xmark & \cmark \\
			Patent~\cite{cheney:2003:integrated} & \xmark & 2 & 1 & Unspecified & \cmark & \cmark & \xmark & \xmark & \xmark \\
			Patent~\cite{yamazaki:1991:method} & \xmark & 4 & 1 & Unspecified & \xmark & \xmark & \xmark & \xmark & \xmark \\
			Patent~\cite{ogrinc:1995:real} & \xmark & 3 & 3 & Unspecified & \xmark & \xmark & \xmark & \xmark & \xmark \\
			Patent~\cite{tian:2014:method} & \xmark & 4 & 2 & Unspecified & \xmark & \xmark & \cmark & \xmark & \xmark \\
			Proposed & \cmark & Any & Any & Any & \xmark & \cmark & \cmark & \cmark & \cmark \\ \bottomrule
	\end{tabular}}
	\caption{Comparison among the proposed architecture and different architectures for simultaneous processing of multiple video signals.}
	\label{tab:comparison}
	\vspace*{-0.1in}
\end{table*}

Generally speaking, simultaneous processing of multiple video sources faces several challenges. First, the frame rate of a video produced from a video source commonly has a small deviation because of the electrical and mechanical restrictions inside the video source hardware~\cite{jack:2011:video}, and this causes a time-varying deviation among the frame rates of the multiple input videos. Second, the video signals generated from multiple sources can not be started at exact time without using a custom external hardware synchronization signal, therefore there is an inevitable delay between the start of the corresponding frames from the sources. Third, when the input analog videos are decoded into appropriate digital formats (such as BT.656~\cite{bt656:2007:interface}) to be suitable for processing, the previous problems appears in the produced digital video signals making them not suitable for simultaneous processing. Finally, Dealing with the massive amount of data provided by the different video sources greatly increases the complexity specially for time critical applications that require fast processing for the input videos.

To implement a system that performs simultaneous processing of multiple video sources, a ready made and direct solution is to use one of the market available video processing hardware cards aided with necessary components to overcome the aforementioned challenges associated with simultaneous processing. Such type of cards are widely used as in~\cite{hashash:2016:high,khalifa:2015:near,khodary:2014:anew,desmouliers:2010:hw,said:2012:embedded,touil:2014:generic,toledo:2007:fpga,pandey:2012:embedded,muller:2014:new}. Another solution is to design and implement a custom video processing platform tailored to specific application such as the patents in~\cite{taylor:1979:video,fu:2010:video,beckwith:1995:video,bennett:2006:video,cheney:2003:integrated,yamazaki:1991:method,ogrinc:1995:real,tian:2014:method}. For example, Guo et al. proposed a multi-video processing board that stitches videos from different sources into one larger video to be displayed on 4K ultra high definition (UHD) display~\cite{guo:2017:fpga}. However, both solutions (the ready made and custom one) are not flexible enough to handle modifications in design requirements of an application such as the desired number, the required video standards, or the frame rates of the input/output videos.

In this paper, we propose a design of a {\archadj} architecture that is suitable for various applications requiring simultaneous processing of multiple video signals. The proposed architecture has several advantages. The proposed architecture is compatible with most video processors, deal with multiple video standards and features, and scalable to process and provide variable number of input and output videos. Moreover, the proposed architecture handles the challenges associated with simultaneous processing and takes into account the mixed-signal considerations, since we deal with both analog and digital video signals, as well as power distribution considerations. In Table~\ref{tab:comparison}, we compare the proposed architecture with different previous architectures for simultaneous processing of multiple video signals in terms of the scalability of the architecture, the number of input and output video signals, the independence of the output video format from the input one, the video standards that the architecture deal with, the ability to process different spatial and temporal resolutions, and the ability to handle the startup delay and the deviation in the temporal resolution. This comparison motivated us to design the proposed architecture as {\archadj} to be used for implementing various applications of simultaneous processing of multiple video signals.

As a case study, we use the proposed architecture for a realization of a video processing system that performs real time visible-near infrared video fusion~\cite{awad:2018:areal}. Academically, this paper provides key guidelines for designing a system that performs simultaneous processing of multiple input videos to be a reference for researchers who intend to design and implement such systems. Upon publication, we make the source files of the hardware design together with the bill of material (BOM) of the case study available online (at the authors website) for use by researchers in the field.

The rest of the paper is organized as follows. We present the problems facing simultaneous multi-video processing systems in Section~\ref{sec:multi_vid_process_problems}. In Section~\ref{sec:prop_arch}, we present the proposed {\archadj} architecture. Section~\ref{sec:design_con} presents the considerations of  analog/digital mixed-signals and power distribution that we take into account in the design of the proposed architecture. In Section~\ref{sec:case_study}, we present a case study that utilizes the proposed {\archadj} architecture for realization of a video processing system that performs real time visible-near infrared video fusion. The experimental results of the implemented case study are presented in Section~\ref{sec:exp_res}. Finally, the paper is concluded in Section~\ref{sec:concl}.

\section{Problems Affecting Simultaneous Multi-Video Processing}
\label{sec:multi_vid_process_problems}

There are two problems facing the simultaneous processing of multiple input video signals. The first problem is the delay between the start of the corresponding frames from the input videos. The second problem is that the frame rate for a video signal is not constant over time which causes a time-varying deviation among the frame rates of the multiple input video signals. When the video signals are decoded, the two problems also affect the produced digital formatted video signals. The mentioned problems are verified experimentally for both analog and digital video signals by using three Swann C510R color cameras~\cite{swann:camera}, a Pioneer DVD-V8000 player~\cite{pioneer:oscilloscope}, a TVP5150 video decoder~\cite{instruments:2013:ultralow}, and a Keysight MSOS054A mixed signal oscilloscope~\cite{keysight:oscilloscope}. The three cameras and the DVD player act as four different and independent analog video sources that provide interlaced NTSC video signals (all with frame rates of approximately 30 fps). We separately discuss the mentioned problems associated with simultaneous multiple video processing in the following subsections.

\begin{figure}[t]
	\begin{minipage}[b]{\linewidth}
		\centering
		\includegraphics[width=0.75\linewidth]{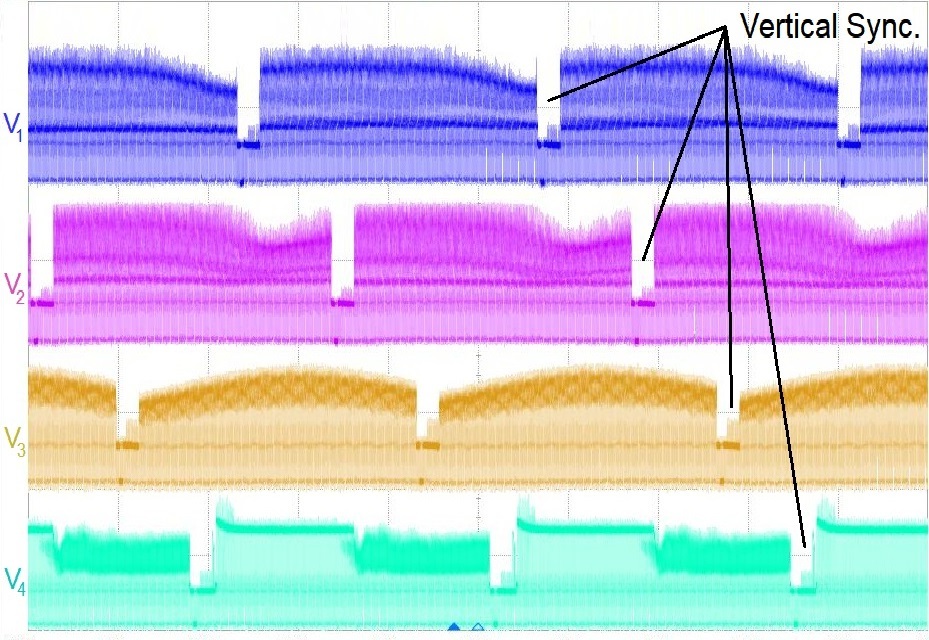}
		\centerline{(a)}\medskip
	\end{minipage}
	\hfill
	\begin{minipage}[b]{\linewidth}
		\centering
		\includegraphics[width=0.75\linewidth]{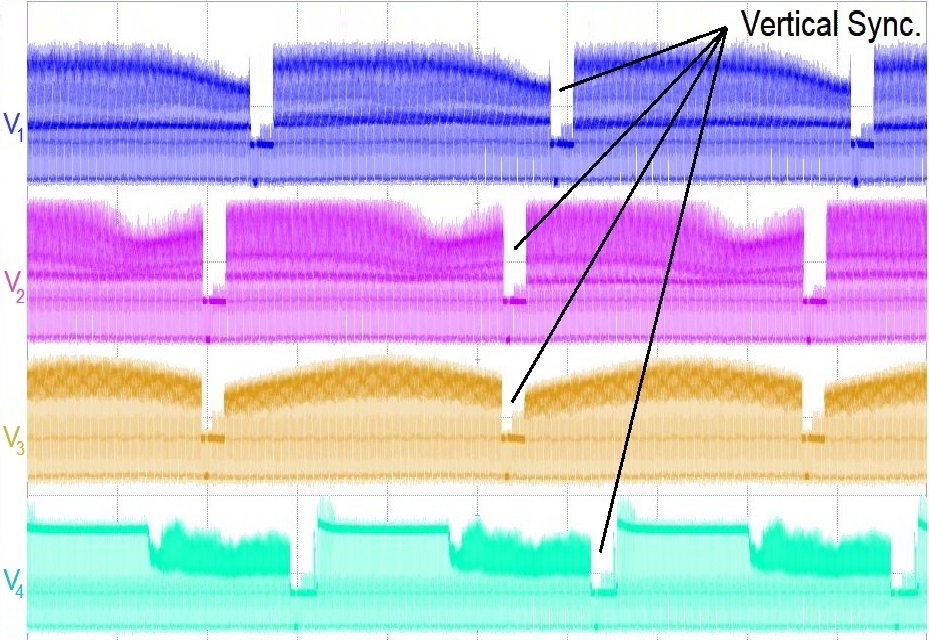}
		\centerline{(b)}\medskip
	\end{minipage}
	\caption{Waveforms of the four input interlaced NTSC video signals. Figures (a) and (b) present waveforms of the same signals captured at different time instances. Each video signal is formed as a stream of fields separated by vertical sync portions.}
	\label{fig:delay_exp}
\end{figure}

\begin{figure}[ht!]
	\begin{minipage}[b]{\linewidth}
		\centering
		\includegraphics[width=0.85\linewidth, frame]{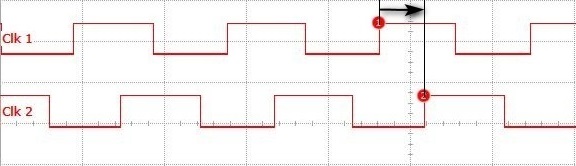}
		\centerline{(a)}\medskip
	\end{minipage}
	\hfill
	\begin{minipage}[b]{\linewidth}
		\centering
		\includegraphics[width=0.85\linewidth, frame]{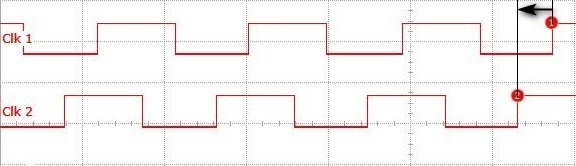}
		\centerline{(b)}\medskip
	\end{minipage}
	\caption{Digital synchronization clocks generated from two TVP5150 video decoders, where each decoder takes a NTSC video signal as an input. Figures (a) and (b) present the synchronization clocks at different time instances.}
	\label{fig:clk_rates_exp1}
\end{figure}

\begin{figure}[ht!]
	\centering
	\includegraphics[width=0.85\linewidth]{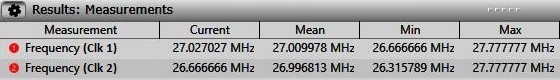}
	\caption{Measured frequencies of the two synchronization clocks in Fig.~\ref{fig:clk_rates_exp1}. The minimum, mean and maximum frequencies are measured using 300 samples for each clock.}
	\label{fig:freq_diff}
\end{figure}

\subsection{Startup Delay}
\label{ssec:delay}

Practically speaking, the start of corresponding frames in videos provided by different sources constitute inevitable delay. The delay occurs as the video signals from different sources can not be started at exact time without using a custom external hardware synchronization signal. We conducted an experiment to verify the delay problem. We powered on the three cameras and the DVD player simultaneously and captured the four analog NTSC video signals using the mixed signal oscilloscope. The waveforms of the four video signals were captured, and we show two different time instance snapshots in Fig.~\ref{fig:delay_exp} (a) and (b). Each video signal is formed as a stream of fields separated by vertical sync portions. As shown in Fig.~\ref{fig:delay_exp} (a), the corresponding frames (fields) of the four video signals do not start together which is clarified form the vertical sync portions shown in the figure.

\begin{figure*}[ht!]
	\centering
	\includegraphics[width=0.8\textwidth]{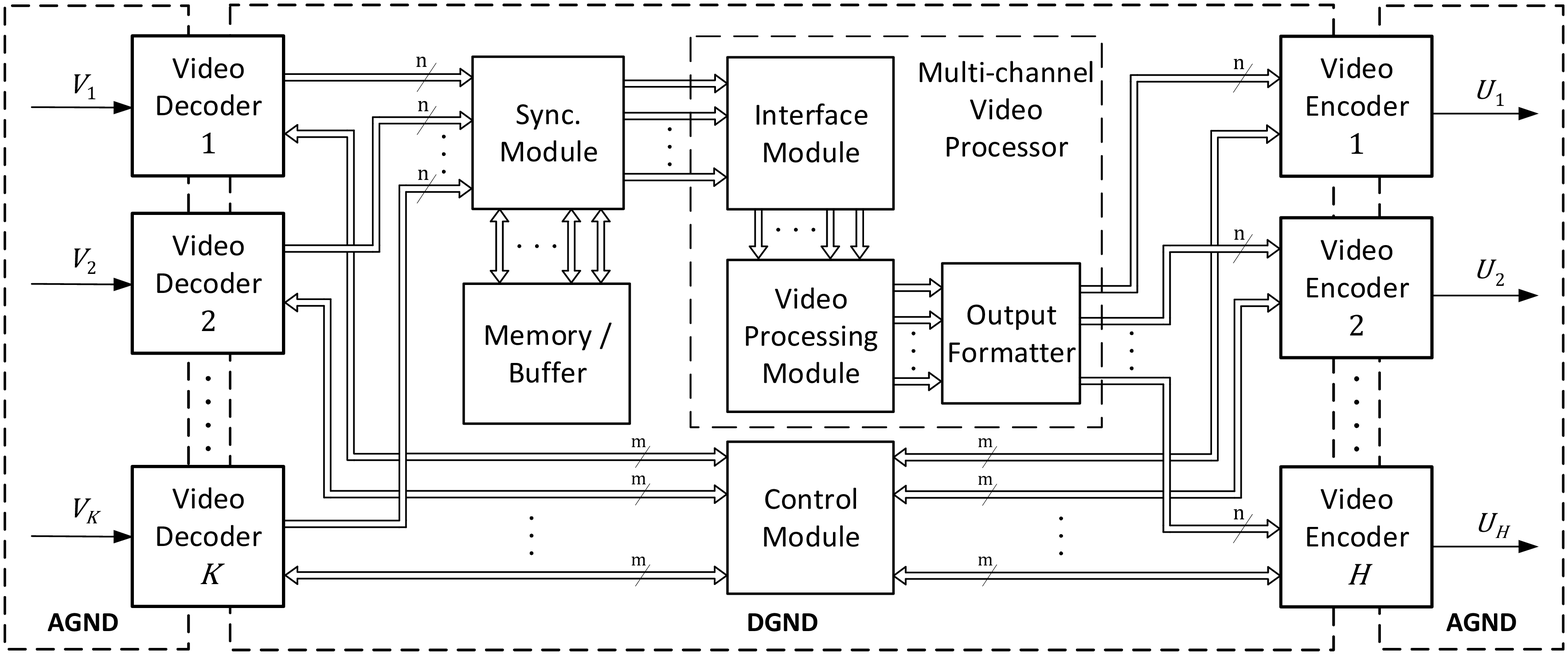}
	\caption{The proposed general architecture of multiple inputs/outputs simultaneous video processing system.}
	\label{fig:general}
\end{figure*}

\subsection{Deviation of Frame Rates}
\label{ssec:diff_rates}

The video generated from a video source has a small variation of frame rate over time due to internal electrical or mechanical hardware limitations of the video source~\cite{jack:2011:video}. This small variation causes a time-varying deviation among the frame rates of the multiple input video signals. As shown in Fig.~\ref{fig:delay_exp} (a) and (b), the time-varying deviation of the frame rates causes the time difference among the starts of corresponding frames/fields in the four video signals to be not constant over time.

\subsection{Deviation of Digital Clock Rates}
\label{ssec:digital_problem}

To be suitable for processing, the input video signals needed to be decoded into corresponding digital formatted video signals. Commonly, a video decoder samples an input along video signal using a synchronization clock into the same number of digital samples for every frame/field such that each sample is synchronized with a clock edge. Because of the time-varying frame rate of the input video signal, the video decoder uses a genlock circuit to accurately tune the rate of the synchronization clock such that the digital samples of a frame/field are produced in the same time interval of the corresponding input frame/filed~\cite{jack:2011:video}. When dealing with multiple input videos, the time-varying synchronization clock for every video poses a leading challenge for simultaneous processing due to the lack of unique clock. Furthermore, the startup delay problem is moved to the decoded digitally formatted signals because the input video signals are not simultaneously started as we indicated previously.

We verify the above problems by decoding two input analog NTSC video signals from two Swann C510R cameras to a corresponding two BT.656~\cite{bt656:2007:interface} digital video signals using two TVP5150 video decoders~\cite{instruments:2013:ultralow}. The video decoder typically produces 900900 samples per frame along with a synchronization clock. We captured the synchronization clocks of the two decoded video signals at two different time instances using the logic analyzer embedded in the mixed signal oscilloscope and plot them in Fig.~\ref{fig:clk_rates_exp1} (a) and (b). We also measure the rates of the two clocks as shown in Fig.~\ref{fig:freq_diff}, which displays the min, max, and mean frequencies of every clock. As shown in the figures, the clock rate of a digital video is not constant over time and there is a clear deviation among the rates of the clocks of the two digital videos.

\section{Proposed {\Archadj} Architecture of Simultaneous Multi-Video Processing}
\label{sec:prop_arch}

In this paper, we are concerned with a simultaneous multi-video processing system that accepts $\numofinputs$ input video signals from different sources and provide processed $\numofoutputs$ output video signals. We propose a {\archadj} architecture that enables a video processor to perform simultaneous processing of the multiple input video signals. The architecture takes the $\numofinputs$ inputs $\{\videoin_1, \dots, \videoin_\numofinputs\}$ from different video sources, converts them to $\numofinputs$ digitally formatted video signals $\{\videoindig_{1}, \dots, \videoindig_{\numofinputs}\}$ each with startup delay $\startupdelay_{i}$ and a time-varying clock rate $\clockrate_{i}(t)$, synchronizes the converted $\numofinputs$ signals to produce $\numofinputs$ synchronized video signals $\{\videoindigsync_{1}, \dots, \videoindigsync_{\numofinputs}\}$ that are ready for processing by a video processor, and encodes the processed video signals into $\numofoutputs$ video outputs $\{\videoout_{1}, \dots, \videoout_{\numofoutputs}\}$. Fig.~\ref{fig:general} shows the proposed general architecture. As shown in the figure, the proposed architecture consists of four main stages: decoding, synchronization, processing, and encoding which are presented next.

\subsection{Video Decoding}
\label{ssec:decoding}

The decoding stage contains all the decoders that are used to convert the input video signals to digitally formated form suitable for digital processing. As shown in Fig.~\ref{fig:general}, the $\numofinputs$ input video signals are passed to corresponding $\numofinputs$ video decoders. Each decoder takes an input video signal $\videoin_i$ with a standard such as NTSC, PAL, SECAM, DVI or HDMI, and produces an appropriate $\numofbits$-bits digitally formated video $\videoindig_i$ in a standard protocol such as BT.656, BT.709, BT.2020~\cite{bt656:2007:interface,nilsson:2015:ultra}, along with a synchronization clock $\clockrate_{i}(t)$. The control module shown in Fig.~\ref{fig:general} configures each video decoder according to the standard of the input video and the required digital format of the converted output video through an interface such as serial bus protocol, e.g., inter-integrated circuit I$^2$C or serial peripheral interface SPI standards.

\subsection{Video Synchronization}
\label{ssec:sync}

The $\numofinputs$-digitally formatted video signals $\videoindig_i$ produced by the decoders are not suitable for simultaneous processing since each video has different startup delay ($\startupdelay_{i}$) and a time-varying clock rate ($\clockrate_{i}(t)$) as we presented in Section~\ref{sec:multi_vid_process_problems}. To synchronize the digital video signals, we adopt a synchronization module that takes one of the digital videos as a reference video and synchronizes the rest $\numofinputs-1$ videos according to the clock rate of the reference video. Without loss of generality, we present the synchronization module below based on the usage of $\videoindig_1$ as the reference video.

\begin{figure}[t]
	\centering
	\includegraphics[width=0.75\linewidth]{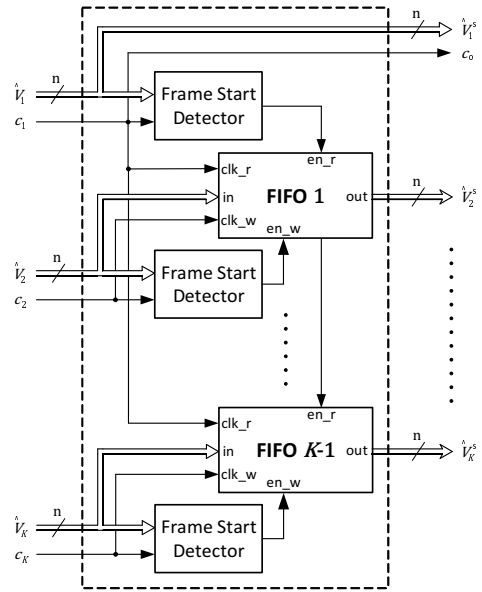}
	\caption{General architecture of the synchronization module. The inputs of the module are $\numofinputs$ different $\numofbits$-bits video signals $\videoindig_{i}$, where $i=1,2,...,\numofinputs$, and their corresponding sampling clocks $\clockrate_{i}(t)$. While, the outputs of the module are $\numofinputs$ synchronized video signals $\videoindigsync_{i}$ and all of them are sampled by the same clock $\clockrate_{o}(t)$.}
	\label{fig:sync_gen}
\end{figure}

As shown in Fig.~\ref{fig:sync_gen}, the synchronization module consists of $\numofinputs$ frame start detectors (FSD) and $\numofinputs-1$ circular first-in-first-out (FIFO) buffers, each with a size equal to the frame size of a digital video signal\footnote{We assume that the frame sizes of all videos are the same}. The $i^{th}$ FIFO buffer holds one frame from the $(i+1)^{th}$ digital video signal such that the first pixel is saved at the first location of the buffer and subsequent pixels are saved in the subsequent locations. A video frame from a digital video signal is written to a FIFO buffer using the clock rate of that digital video signal such that one pixel is written per clock. The writing of a frame is triggered by the frame start which is detected by the corresponding FSD. To ensure that the pixels at the same spatial positions output from the buffers at the same time, the read operation is performed with same clock rate $\clockrate_{o}(t)$ which equal to the clock rate of the reference video ($\clockrate_1(t)$ in this case as we use $\videoindig_1$ as the reference video). Additionally, the read operation is started simultaneously across all buffers at the frame start of the reference video to make sure that the synchronized signals have zero delays.

Note that, the synchronization module guarantees that individual pixels in the same spatial positions in all buffers are delivered at the same time to the processing module. However, these pixels are not necessary related to the same temporal frame. For instance, if $\clockrate_{o}(t)$ is the slowest clock rate among the clock rates of the $\numofinputs$ videos, i.e., $\clockrate_{o}(t)=\mathbf{\mathrm{min}}(\clockrate_{i}(t);\forall{i\in\{1,\dots,\numofinputs\}})$, then individual pixels in same spatial position in all buffers are correspond to future frames with respect to the corresponding pixel (in the same spatial position) in the buffer of the reference video. This temporal difference between pixels does not have any impact on the desired process of the proposed architecture because of the temporal redundancy nature of the video signals. Specifically, the current frame has a lot in common with the next and former frames, so pixels at same spatial positions across few temporal frames of a video are approximately similar.

\subsection{Video Processing}
\label{ssec:processing}

In the processing stage, the video processor is used to perform $\numofoutputs$ desired functions on the $\numofinputs$ synchronized digitally formatted video signals $\videoindigsync_i$ and produce $\numofoutputs$ processed digital output videos. In general, the video processor contains three main modules: interfacing, processing and output formatter modules, as shown in Fig.~\ref{fig:general}. The interface module decodes the digital video signal into raw video data consisting of the pixel values inside each frame along with the embedded horizontal and vertical synchronization signals of the video. For example, the BT.656 digitally formated video consists of raw 4:2:2 YCbCr values with embedded horizontal and vertical sync. The video processing module represents mainly the core of the system, and it carries out the implementation of the desired video processing application on the input pixel values. The output formatter is responsible for reformatting the processed raw video data into a digital format, such as BT.656.

\subsection{Video Encoding}
\label{ssec:encoding}

The encoding stage contains all the encoders that are used to convert the processed digital videos to appropriate video format in order to be displayed on the desired displays as intended. As shown in Fig.~\ref{fig:general}, the $\numofoutputs$-processed digital formatted video signals are passed to $\numofoutputs$-video encoders. The clock needed to encode the video outputs is the temporal clock $\clockrate_{o}(t)$ selected from the synchronization module. The control module shown in Fig.~\ref{fig:general} configures each video encoder according to the standard protocol of the input digitally formated video and the required format of the converted output video through an interface such as serial bus protocol, e.g., inter-integrated circuit I$^2$C or serial peripheral interface SPI standards.

\begin{figure*}[ht!]
	\centering
	\includegraphics[width=.75\textwidth]{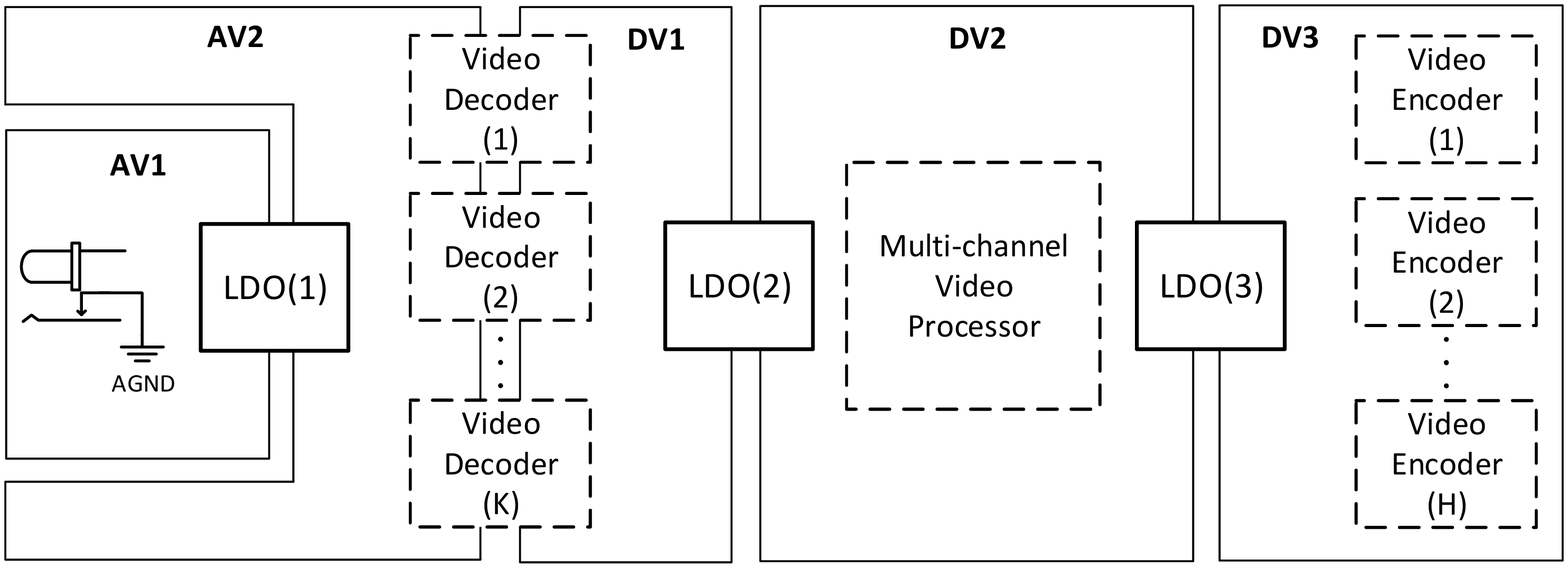}
	\caption{Power planes distribution in the proposed architecture with-respect-to the video decoders, encoders and the three low-dropout (LDO) voltage regulators required for making the different voltages and power sources stable. The LDO voltage regulator ICs are well placed to avoid crossing a power trace between two different power planes.}
	\label{fig:pow_gen}
\end{figure*}

\section{Design Considerations of The Proposed Architecture}
\label{sec:design_con}

\subsection{Mixed-Signal Design Considerations}
\label{ssec:mixed_sig}

The proposed {\archadj} architecture in Fig.~\ref{fig:general} shows the existence of sensitive analog signals at the inputs of the decoders, $\videoin_i$, and the outputs of the encoders, $\videoout_i$, in addition to the existence of digital signals, such as $\videoindig_i$ and $\videoindigsync_i$, between the mentioned modules and the video processor. Thus, the proposed architecture is considered as a mixed-signal system. There are two main issues hindering performance of the mixed-signal systems. First, the analog signals are highly susceptible to transient noise due to electromagnetic interference (EMI) and crosstalk. Such noise is caused by the large number of traces with different mechanical dimensions covering large area, and thus might form a considerable antenna effect~\cite{eastman:1996:considerations}. The electromagnetic field passing over a trace causes undesired stimulated electrical current to flow in that trace. Second, digital signals are noisy due to fast clock switching and therefore it may inject considerable noise into the other analog signals on the proposed architecture.

In order to design our architecture to comply with mixed-signal considerations and to maintain high video quality, we took into account two main principles of electromagnetic compatibility (EMC) and EMI.
\begin{enumerate}
	\item Currents should be returned to their sources over the minimal possible loop area as locally and compactly as possible, otherwise a loop antenna will be created~\cite{ott:2001:partitioning}.
	\item Only one reference plane should be included in the system, otherwise a dipole antenna will be created as a result of using two references~\cite{ott:2001:partitioning}.
\end{enumerate}
We make our system less susceptible to EMI by reducing the coupling and using decoupling filter capacitors. Additionally, proper grounding is used to divert EMI away from the analog signals by providing low-impedance path. Furthermore, we avoid parasitic antenna structures by reducing loops of circulating currents.

Taking into consideration the aforementioned EMC and EMI principles, we solved some of the mixed-signal issues using the ground plane isolation. A ground plane serves as the return path for currents from the different modules besides reducing the EMI and crosstalk electrical noise. As shown in Fig.~\ref{fig:general}, we split the ground plane into two separated planes one for the analog signals ground (AGND) and the other for the digital signals ground (DGND) to keep the analog part away from the digital switching effects. Splitting the ground plane was not an easy choice because routing a trace over the split would greatly increase the interference and crosstalk noises~\cite{ott:2001:partitioning}. Thus, modules placement and partitioning were performed efficiently to avoid routing traces from a ground plane to another as shown in Fig.~\ref{fig:general}. The design modularity facilitated partitioning of the proposed architecture. Thus, the architecture was divided into two parts. One contains all the traces that carry analog signals and transfer them to and from the decoders and encoders ICs. Similarly, the other part contains the connectors of the video processor as well as all the traces that carry digital signals and transfer them to and from the decoders and encoders ICs. The only way to transfer signals between the analog and digital planes is through the mixed signals decoders and encoders ICs, not over the split between the two different planes. According to the aforementioned EMC and EMI principles, we tied the two analog and digital ground planes together at a single tie point for each decoder IC, to avoid generating a dipole antenna due to the use of two ground planes.

\subsection{Power Distribution Considerations}
\label{ssec:power_dist}

ICs of the video decoders and encoders require mostly one or more of the three standard voltage levels which are 1.8, 3.3 and 5 DC Volts either for analog or digital parts. The analog DC voltages can be driven from an external DC power supply, while the digital DC voltages can be supplied from the video processor power pins. A noisy power supply may considerably affect the performance of the system because power is distributed everywhere, thus power source noise will propagate all over the circuit. Such noise source may be the external power supply noise or the degradation of power driven from different sources. In the design of the proposed architecture, we isolated and reduced the effect of these possible noises using appropriate strategies such as using voltage regulation, ferrite beads and power planes. Such strategies enabled us to handle the errors that may occur due to any of the aforementioned noise sources during system operation.

The power lines/planes are responsible for transferring power from the external power sources to the load devices. Each power line/plane contains noise in the form of high frequency EMI and undesired power spikes. Thus, we used a ferrite bead for each power line/plane in order to isolate these noises from reaching the loaded devices. Each ferrite bead is used with a couple of bypass capacitors to form a high frequency suppression filter that reduces EMI. Additionally, the ferrite beads withstand the sudden changes in the input current and the unwanted power spikes and transients.

We used a low-dropout (LDO) linear voltage regulator at each voltage source because the ICs, in general, are very sensitive to any supply voltage variations. LDO is a DC linear voltage regulator that can regulate the output voltage even if the supply voltage is very close to the output voltage. The performance of the ICs may be affected if the supply voltage variation exceeds certain limits according to the datasheets of the ICs. Furthermore, the output of a voltage source is not ideally stable due to: (1) environmental conditions and (2) interference with other different signals~\cite{brooks:2000:splitting} that causes an additive or subtractive noise to the voltage level. An LDO voltage regulator provides a constant DC voltage level independent on the amount of load current, so that any change in the output of the voltage sources will be isolated from the voltage supply pins of the ICs. The dropout voltage of a regulator is the minimum difference between the input and output voltages, therefore we preferred LDO over normal voltage regulators. As shown in Fig.~\ref{fig:pow_gen}, at least three LDO voltage regulator ICs are required for the proposed architecture. LDO(1) takes the analog domain input DC voltage (AV1) from external DC power supply and outputs regulated DC voltage levels (AV2) necessary for the video decoders within the required supply voltage variations. While, LDO(2) and LDO(3) take digital domain DC voltage (DV2) from the video processor and provide DC voltage levels (DV1) and (DV3) required for the video decoders and encoders, respectively. The required number of LDOs is determined according to the maximum required load current $I_{\text{max}}$ sufficient such that all the load devices connected to each LDO work properly, and can formulated as
\begin{equation}
I_{\text{max}} = I_{1} + I_{2} + ... + I_{L},
\label{eq:ldo_num}
\end{equation}
where $L$ is the number of load devices connected to an LDO and $I_{1}$, $I_{2}$, ... , $I_{L}$ are the load currents for each device requiring the LDO. If $I_{\text{max}}$ exceeds the maximum current derive capability of the LDO, additional LDOs should be added accordingly until the total output current covers $I_{\text{max}}$.

For the supply voltage levels required to derive the loads properly, we used split power planes, as shown in Fig.~\ref{fig:pow_gen}. This is because using thin traces to transfer power between the sources and different ICs in the architecture may cause voltage drop in the levels of the power signals due to the trace equivalent resistivity, because the area of a trace carrying a signal is inversely proportional to the impedance of that trace. In other words, a trace will resist the power that it carries, which in turn may affect performance of the load devices that expect certain power level delivered from the LDO voltage regulators~\cite{brooks:2000:splitting}. The area of each used power plane was as maximum as possible to provide a uniform distribution for all power signals and hence decrease the chance of signal crossing to facilitate the routing between different modules. Fig.~\ref{fig:pow_gen} shows how we carefully partitioned the proposed architecture in the case study that will be presented in Section~\ref{sec:case_study}. As shown in Fig.~\ref{fig:pow_gen}, we carefully placed the different LDO voltage regulator modules with-respect-to the video decoders, encoders and video processor in order to avoid overlapping two different power planes or having a trace crossing the split between two different planes. There exist another great advantage of having power planes at a dedicated layer(s) as well as the ground planes at another dedicated layer(s), based on the process used for fabrication. This advantage is that any two facing areas of copper pours create a large parallel plate decoupling capacitor that prevents noise from being coupled from one circuit to another through the power supply. Even though this parasitic capacitor might impact the signal bandwidth, but it is acceptable for such application.

\section{Case-Study: Fusion Based Video Enhancement}
\label{sec:case_study}

We present in this paper a case study that utilizes the proposed {\archadj} architecture for realization of a real time visible-near infrared video fusion approach~\cite{awad:2018:areal}\footnote{The FPGA implementation of the fusion approach~\cite{awad:2018:areal} is proposed by some of the authors of this paper. Therefore, this paper brings the previous work to real-time realization. We modified the FPGA implementation in~\cite{awad:2018:areal} to be suitable for video fusion, because the original implementation was proposed for image fusion only. Details of the design and the FPGA implementation of the video fusion is a subject of another article contribution.}. The fusion approach adaptively injects complementary details from a near infrared (NIR) frame into the visible (VS) one for the same scene in order to compensate for missing details and obtain an enhanced visible image. In the following subsections, we present detailed architecture and design considerations of the implemented case study.

\subsection{Case-Study Architecture}
\label{ssec:case_arch}

\begin{figure*}[ht!]
	\centering
	\includegraphics[width=0.75\textwidth]{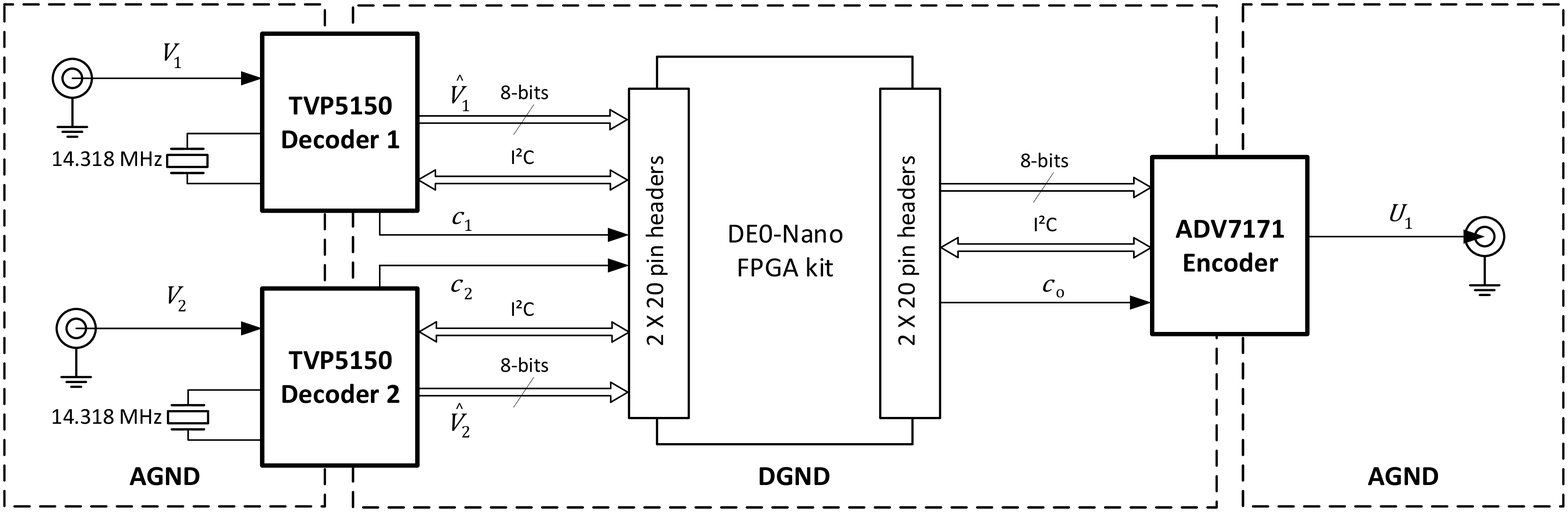}
	\caption{Overall architecture of the case-study implemented using the proposed video processing general architecture. The two input video signals are entered via BNC connectors and converted from NTSC/PAL analog form to BT.656 8-bits digital video steams using two TVP5150 decoders. The connected FPGA device then performs the required processing on the digital video streams and provide the result to be converted into analog NTSC/PAL form using ADV7171 encoder and the output is exited via a BNC connector. The programming of decoders and encoders is done by the FPGA device using I$^2$C serial interfaces.}
	\label{fig:arch}
\end{figure*}

\begin{figure}[ht!]
	\centering
	\includegraphics[width=.75\linewidth]{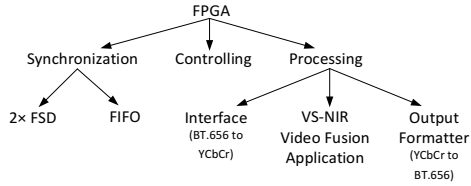}
	\caption{Tree diagram of the main tasks running on the FPGA device that is used in the case-study.}
	\label{fig:tree}
\end{figure}

The main design requirements that we need to fulfill for our case study are two video inputs ($\numofinputs=2$), one video output ($\numofoutputs=1$), three main DC supply voltage levels of 1.8, 3.3 and 5 Volts for power requirements, a video processor, and programmable decoder/encoder ICs. We present the case study design in Fig.~\ref{fig:arch}. The inputs are two analog NTSC/PAL video signals connected via two BNC sockets and the output is an analog NTSC/PAL video signal. In this case study, we used an FPGA device as a video processor in addition to implement the synchronization and control modules shown in Fig.~\ref{fig:general}. Specifically, the FPGA is used for (a) implementing the synchronization module (presented in Sec.~\ref{ssec:sync}), (b) controlling the decoder/encoder ICs, and (c) performing the actual processing (which is the video fusion in this case). We conclude the main tasks that run on the FPGA device in Fig.~\ref{fig:tree}. In the following, we separately illustrate the main stages of the implemented case study.

\subsubsection{Video Decoding}
\label{sssec:case_decoding}

We chose two TVP5150 low power video decoder ICs from Texas Instruments~\cite{instruments:2013:ultralow} to convert the NTSC/PAL analog video signals into ITU-R BT.656 digital video format that consists of 8-bits 4:2:2 luminance-chrominance (YCbCr)~\cite{itu:2007:bt656}. The decoders require an external 14.31818 MHz crystal oscillator (AS-14.31818-18-SMD) to generate the required sampling clock as shown in Fig.~\ref{fig:arch}.

\begin{figure}[t]
	\centering
	\includegraphics[width=0.75\linewidth]{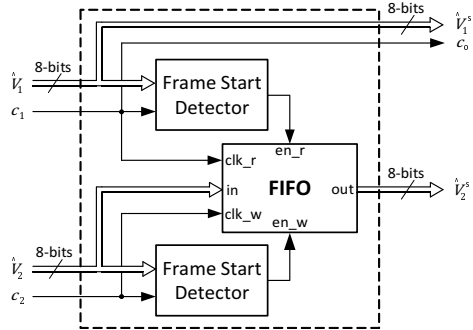}
	\caption{Block diagram of the synchronization module. The inputs of the module are two different 8-bits video streams, $\videoindig_{1}$ and $\videoindig_{2}$, and their corresponding synchronization clocks, $\clockrate_{1}$ and $\clockrate_{2}$. While, the outputs of the module are two synchronized video streams, $\videoindigsync_{1}$ and $\videoindigsync_{2}$, and one synchronization clock $\clockrate_{o}$ for both synchronized video streams.}
	\label{fig:sync}
\end{figure}

\begin{figure}[t]
	\centering
	\includegraphics[width=0.8\linewidth]{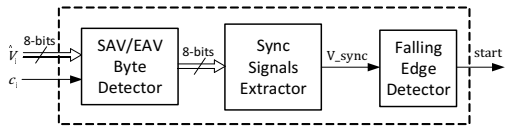}
	\caption{Block diagram of the frame start detector (FSD) used in the synchronization module shown in Fig.~\ref{fig:sync}.}
	\label{fig:fsd}
\end{figure}

\begin{figure*}[t]
	\centering
	\includegraphics[width=.75\textwidth]{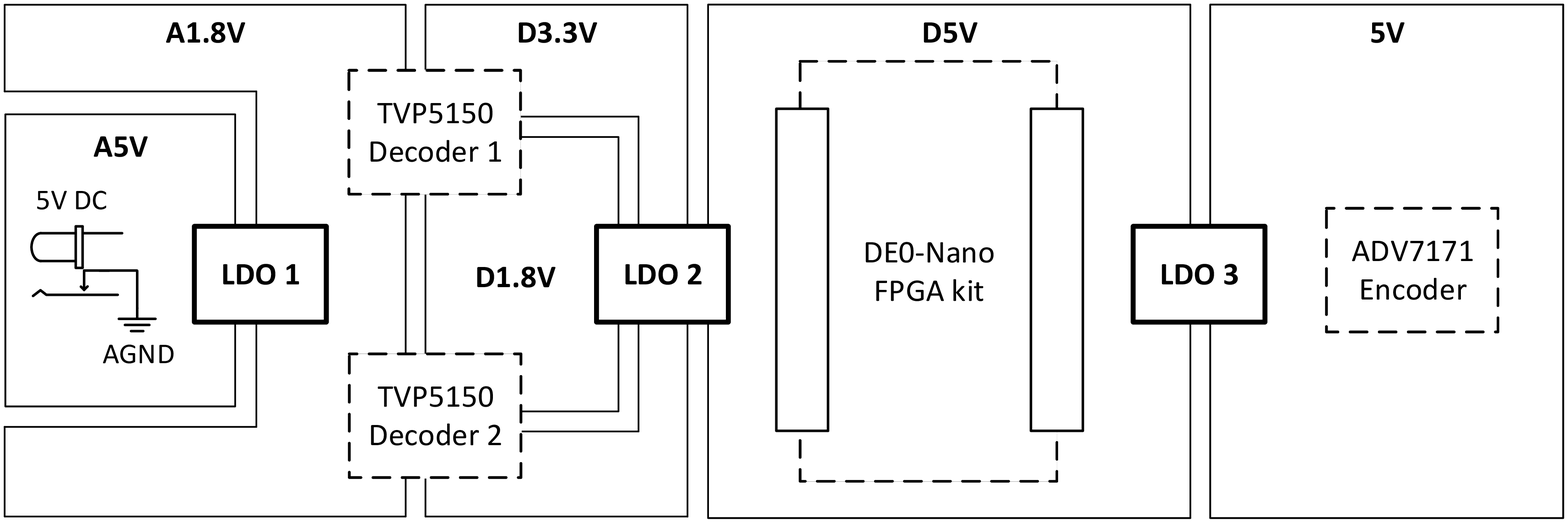}
	\caption{Power planes distribution in the implemented case-study. The LDO voltage regulator ICs are well placed to avoid crossing a power trace between two different power planes.}
	\label{fig:power}
\end{figure*}

\subsubsection{Video Synchronization}
\label{sssec:case_sync}

The two digital video streams provided by the decoders are synchronized using the synchronization module presented in Sec.~\ref{ssec:sync}. We implement the synchronization module using an FPGA device. The synchronization module consists of three submodules: two frame start detectors (FSD) and a single circular first-in-first-out (FIFO) buffer of size $720\times{480}$ bytes (the size of one BT.656 frame), as shown in Fig.~\ref{fig:sync}. An FSD is used to detect beginning of every frame from the input digital stream. In the case study, we selected $\videoindig_{1}$ as the reference video i.e, $\clockrate_{o}(t)=\clockrate_{1}(t)$ and we buffer $\videoindig_{2}$. As shown in Fig.~\ref{fig:sync}, the FSD of $\videoindig_{2}$ triggers the start of the writing operation to ensure that first location in the FIFO contains the first pixel followed by the sequential pixels of the current frame from $\videoindig_{2}$. The FSD of $\videoindig_{1}$ triggers the start of the reading operations to ensure that the first output pixel from the FIFO corresponds to the first pixel of the current frame from $\videoindig_{2}$. This provides a new version of $\videoindig_{2}$ but synchronized with $\videoindig_{1}$ and both are with the same clock rate of $\clockrate_{o}(t)$.

The FSD is presented in Fig.~\ref{fig:fsd}. The FSD first detects the start of active video (SAV) byte that occurs in each line of the video at the start of the active pixels in the line. This byte contains horizontal sync (H$\_$sync), vertical sync (V$\_$sync) and field ID (F$\_$id) indicator bits. The V$\_$sync bit has a value of 0 when the upcoming bytes in the stream contain active pixels and has a value of 1 when the upcoming bytes represent vertical blanking. Therefore, a 1 to 0 transition in the V$\_$sync bit indicates a start of a new frame. To detect the start of a frame, the V$\_$sync bit is extracted from the SAV byte using a V-sync-extractor block and passed to a falling-edge-detector block to detect the start of a frame.

\subsubsection{Video Processing}
\label{sssec:case_processing}

In order to perform the video fusion application, we selected the video processor to be an FPGA device because FPGA devices are suitable for embedded video processing systems as they consist of registers and computational logic units that can perform parallel, concurrent and pipelined operations~\cite{gokhale:2006:reconfigurable}. The used FPGA device in the case study is the Altera Cyclone IV EP4CE22F17C6 FPGA device~\cite{terasic:2012:de0} mounted on the DE0-Nano board. The FPGA device also is responsible for controlling the functionality of the decoders and the encoder using I$^2$C serial interface as shown in Fig.~\ref{fig:arch}. Note that, in our case study, the decoded digital video signals are passed to the FPGA using two connectors. Although the connectors are compatible with the used FPGA kit, any other kits can be connected to the proposed implementation using different connectors.

\subsubsection{Video Encoding}
\label{sssec:case_encoding}

We chose the ADV7171 video encoder IC~\cite{devices:2008:digital} to convert the digital output from the FPGA device from ITU-R BT.656 8-bits 4:2:2 YCbCr format to an NTSC/PAL analog video signal in order to be displayed on a screen via a BNC connector as shown in Fig.~\ref{fig:arch}. The ADV7171 encoder also modulates the luminance and chrominance (Y/C) components to generate the composite video signal. To operate properly, the encoder is fed by the reference clock that was selected by the synchronization module ($\clockrate_{o}(t)$).

\subsection{Design Considerations of The Case-Study}
\label{ssec:case_des}

\begin{figure*}[ht!]
	\centering
	\includegraphics[width=.75\textwidth]{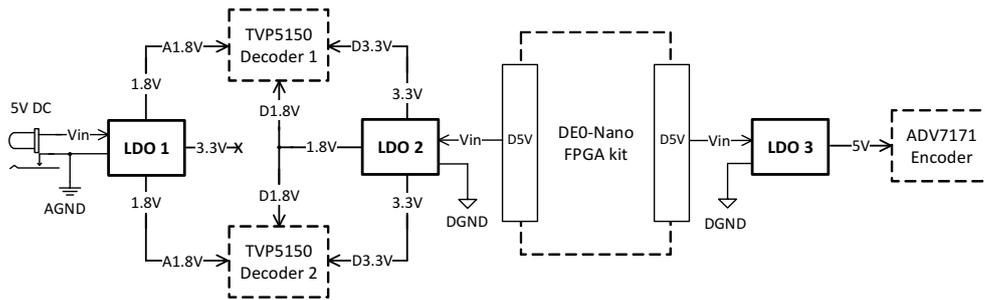}
	\caption{The three low-dropout (LDO) voltage regulators used for making the different voltages and power sources stable. LDO 1 accepts analog 5 V from the external power supply through a barrel jack connector and provides a fixed analog 1.8 V to the two decoders. LDO 2 accepts digital 5 V from the FPGA board through the pin-headers connector and provides fixed digital 1.8 and 3.3 V to the two decoders. Similarly, LDO 3 accepts digital 5 V, but it provides stable 5 V to the encoder.}
	\label{fig:regulators}
\end{figure*}

In the implementation of our case-study, each decoder requires three different voltage levels which are analog domain DC 1.8 V supplied via the analog power plane and digital domain DC 1.8 and 3.3 V supplied via the digital power plane. The encoder requires only digital domain 5 V DC voltage. Fig.~\ref{fig:power} shows the split power planes for the required analog and digital domains voltage levels for the used ICs as recommended in Section~\ref{sec:design_con}. An external DC power supply is used to provide the analog domain DC voltage, while the digital domain DC voltages are supplied from the 5 V DC output pin of the FPGA kit as shown in Fig.~\ref{fig:regulators}. We choose two AP7312-1833W6-7 LDO voltage regulators~\cite{diodes:2011:dual}, for LDO 1 and LDO 2, with 5 V input and two regulated outputs of 1.8 and 3.3 V as shown in Fig.~\ref{fig:power} and Fig.~\ref{fig:regulators}. We also choose an MCP1700T-5002T/TT three terminal LDO voltage regulator~\cite{microchip:2013:low} for LDO 3 with 5 V input and 5 V regulated output. Besides achieving the modularity, we selected the aforementioned LDO voltage regulators to accomplish scalability because the maximum current derive capability for the AP7312 and MCP1700T LDOs are 150 and 250 mA, respectively. On the other side, the maximum load current of the used decoders and encoder ICs are 32.9 and 37 mA, respectively~\cite{instruments:2013:ultralow,devices:2008:digital}. Thus using~\eqref{eq:ldo_num}, a single AP7312 LDO can source current to around 4 encoders, while a single MCP1700T LDO can source current to around 6 decoders. 

\begin{figure}[t]
	\begin{minipage}[t]{\linewidth}
		\centering
		\includegraphics[width=0.85\linewidth]{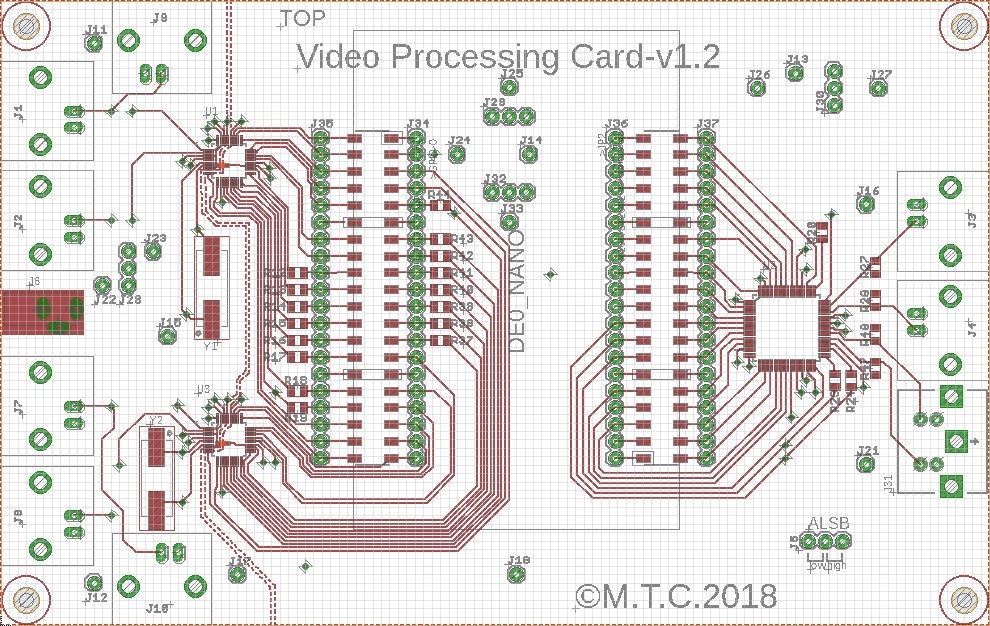}
		\centerline{(a)}\medskip
	\end{minipage}
	\hfill
	\begin{minipage}[t]{\linewidth}
		\centering
		\includegraphics[width=0.85\linewidth]{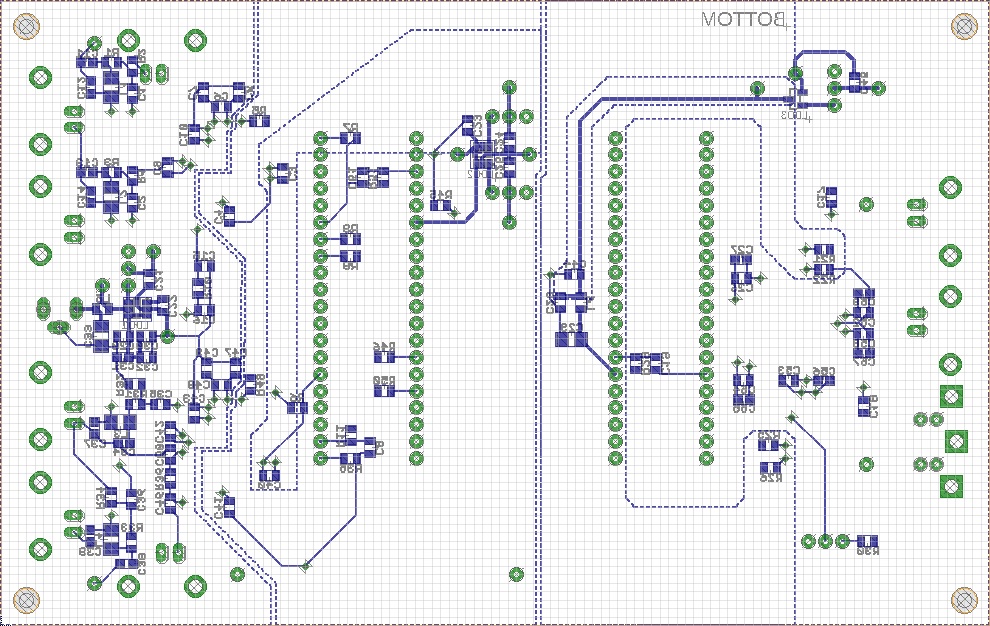}
		\centerline{(b)}\medskip
	\end{minipage}
	\caption{Printed-circuit-board (PCB) (a) top and (b) bottom layers of the implemented dual-channel video processing system.}
	\label{fig:layers}
\end{figure}

In order to achieve complete signal isolation, the 14.31818 MHz crystal oscillators and its associated traces are kept with acceptable clearance of eight to ten times the associated trace width from any adjacent analog signals specially the analog input video signals. This clearance is employed to prevent the high frequency digital switching clock from affecting the analog signals and it is about ten times the width of the used traces~\cite{instruments:2013:ultralow}. We used two crystal oscillators (AS-14.31818-18-SMD) instead of only one to make the board partitioning easier, besides using of only one crystal oscillator may cause clock skew between the two video decoders.

\begin{figure}[t]
	\centering
	\includegraphics[width=0.85\linewidth]{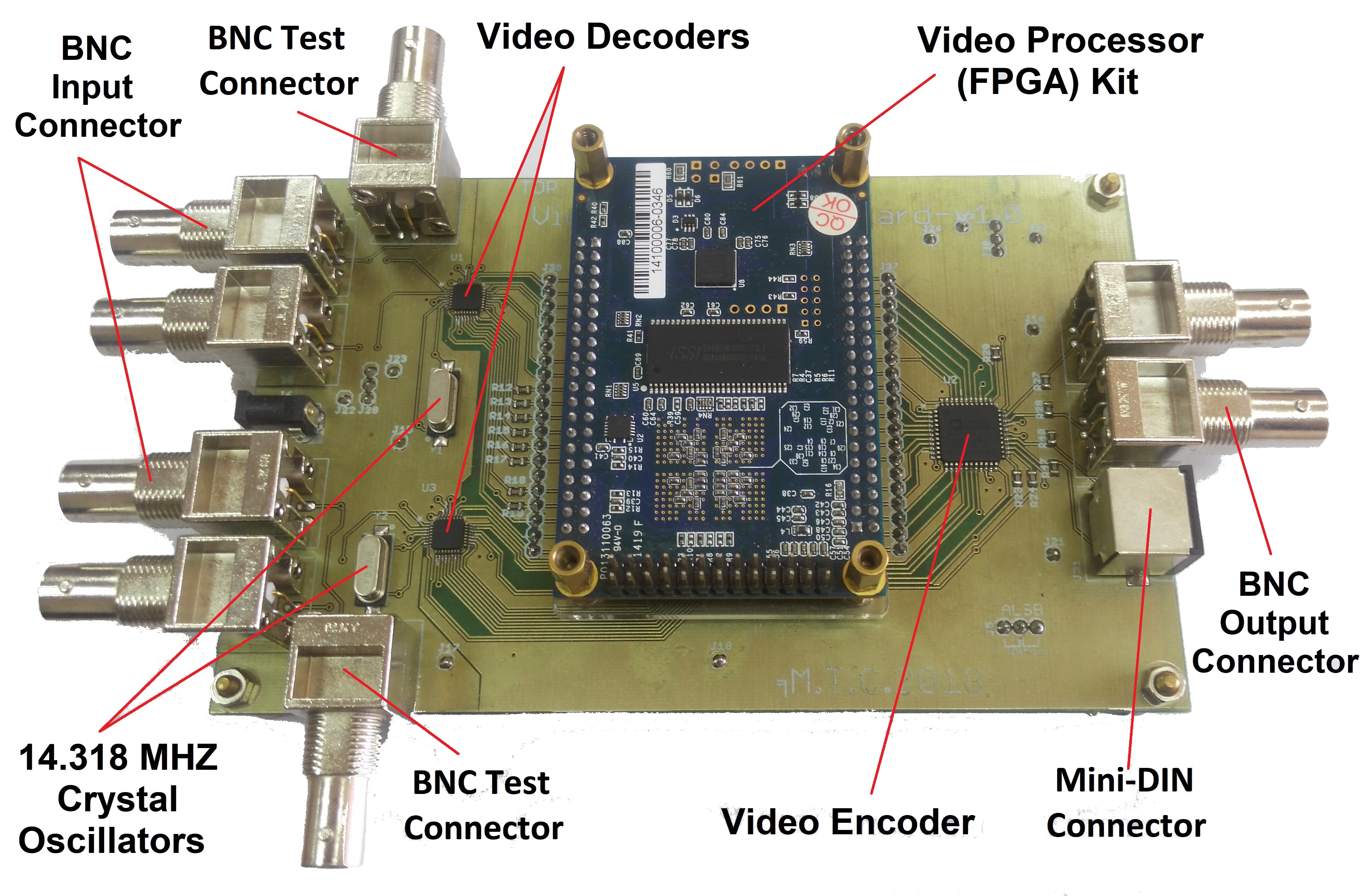}
	\caption{Structure of the implemented dual-channel video processing system.}
	\label{fig:plat_str}
\end{figure}

\section{Experimental Results}
\label{sec:exp_res}

\begin{figure}[ht!]
	\centering
	\includegraphics[width=0.85\linewidth]{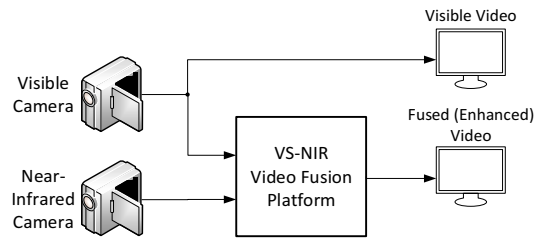}
	\caption{Setup of the dual-channel real-time video processing system.}
	\label{fig:sys_setup}
\end{figure}

\begin{figure}[ht!]
	\centering
	\includegraphics[width=0.85\linewidth]{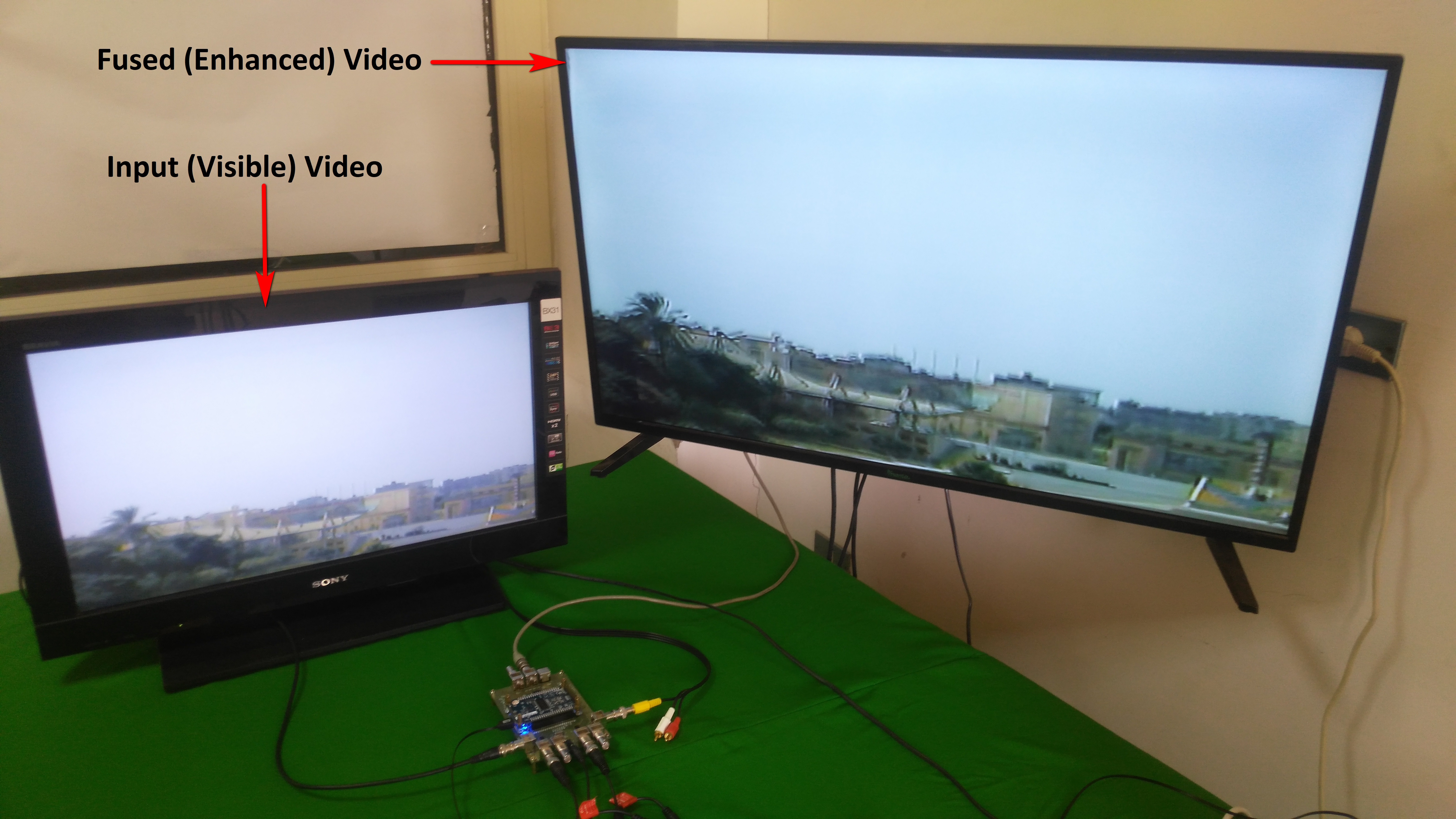}
	\caption{The complete dual-channel real-time video processing system that performs the implementation in~\cite{awad:2018:areal} of the  VS-NIR fusion based enhancement approach in~\cite{elliethy:2017:fast}.}
	\label{fig:plat_sys}
\end{figure}

In this section, we present the realization of the fusion based video enhancement system (case study) presented in Sec.~\ref{sec:case_study}. We use two analog cameras to feed the fusion system with the visible and near-infrared video signals. Two monitors are used to view the input visible video and the output fused enhanced video as shown in Fig.~\ref{fig:sys_setup} and Fig.~\ref{fig:plat_sys}. We design the schematics and layout of the realization of our case study using AUTODESK EAGLE PCB design software~\cite{eagle:pcb}. Printed-circuit-board (PCB) of the proposed system was implemented using the following specifications: fill material of FR-4 TGBO, dimension of (14.9 cm $\times$ 9.4 cm), thickness of 1.6 mm, copper thickness of 1 oz, qualification of 1pc class2-A600 with no special fabrication requirements, and with two copper layers containing all the video decoders, encoders, voltage regulators, input/output and video processor kit connectors along with all other supplementary components as shown in Fig.~\ref{fig:layers}. In Fig.~\ref{fig:plat_str}, we present the implemented system integrated with the DE0-Nano~\cite{terasic:2012:de0} FPGA kit. 

The output video was provided in real-time at frame rate of 30 fps for standard NTSC input videos. Table~\ref{tab:pow_cons_dev} presents the average power consumed by the main devices of the proposed system: two AP7312 LDO regulators, two TVP5150 video decoders, an MCP1700 LDO regulator, and an ADV7171 video encoder. In Table~\ref{tab:pow_cons_sup}, we present the average power consumption measured at both analog and digital voltage supplies to the proposed system.

\begin{table}[t]
	\centering
	\begin{tabular}{|l|c|c|}
		\hline
		& Power Consumption & Total \\ \hline
		$2\times$ AP7312 & 45 mW & \multirow{4}{*}{1.5695 W} \\ \cline{1-2}
		$2\times$ TVP5150 & 230 mW &  \\ \cline{1-2}
		MCP1700 & 44.5 mW & \\ \cline{1-2}
		ADV7171 & 1.25 W &  \\ \hline
	\end{tabular}
	\caption{Power consumption values for the main devices of the implemented system according to their corresponding datasheets.}
	\label{tab:pow_cons_dev}
\end{table}

\begin{table}[t]
	\centering
	\begin{tabular}{|l|c|c|}
		\hline
		& Power Consumption & Total \\ \hline
		Analog DC Supply & 350 mW & \multirow{2}{*}{1.85 W} \\ \cline{1-2}
		Digital DC Supply & 1.5 W &  \\ \hline
	\end{tabular}
	\caption{Measured power consumption values for the input voltage supplies of the implemented system.}
	\label{tab:pow_cons_sup}
\end{table}

\section{Conclusion}
\label{sec:concl}

In this paper, we propose a {\archadj} architecture that enables a video processor to perform simultaneous processing of multiple input video signals. The architecture takes $\numofinputs$ inputs from multiple video sources, converts them to $\numofinputs$ digitally formatted video signals each with different startup delay and time-varying clock rate, synchronizes the converted $\numofinputs$ digital video signals to be ready for processing by a video processor, and encodes the processed video signals into $\numofoutputs$ video outputs. The proposed architecture has many advantages of being compatible with most video processors, modular, upgradeable to deal with multiple video standards and features, can be used for various video processing implementations, scalable to process and provide variable number of input and output videos, and takes into account all mixed-signal and power distribution considerations. As a case study of the proposed architecture, we utilize the proposed architecture for a realization of a real time video fusion system that combines near infrared and visible video sources to produce an enhanced video. The fusion system produces  output video in real-time at frame rate of 30 fps for standard NTSC input videos.

\bibliographystyle{spphys}
\bibliography{jnl_abbrev,IEEEabrv,refs}

\end{document}